\documentclass{Rinton-P9x6}
\usepackage{latexsym}
\usepackage{amssymb}
\newcommand{\pv}{\phi_{\mathrm V}}
\def\nn{\nonumber\\ }
\def\6{\partial }
\def\.{\dot }
\def\5{\bar } 
\def\4{\frac{1}{4}}
\def\2{\frac{1}{2}}
\def\a{\alpha}

\def\c{\chi}
\def\d{\delta}
\def\e{\epsilon}           
\def\f{\phi}               
\def\g{\gamma}

\def\j{\psi}

\def\m{\mu}

\def\o{\omega}
\def\q{\theta}                    
\def\r{\rho}                      
\def\s{\sigma}                    

\def\vf{\varphi}
\newcommand{\pa}{\partial}
\newcommand{\D}{\Delta}

\renewcommand{\ll}{\langle}
\newcommand{\rr}{\rangle}
\begin{document}
\title{The Casimir effect for susy solitons\footnote{{\scriptsize\rm To appear
in the Proceedings to the 6th Workshop on Quantum Field Theory under the Influence of External Conditions (QFEXT03)
Norman, Oklahoma, USA, 2003/09/15-19}}} 

\author{Anton Rebhan}

\address{Institute for Theoretical Physics, Vienna
University of Technology,\\
Wiedner Hauptstr.~8--10, A-1040 Vienna, Austria}

\author{Peter van Nieuwenhuizen}

\address{C.~N.~Yang Institute for Theoretical Physics,\\
SUNY Stony Brook, NY 11794-3840 USA}

\author{Robert Wimmer}

\address{Institute for Theoretical Physics, University of Hanover,\\
Appelstr.~2, D-30167 Hanover, Germany}

\maketitle

\begin{picture}(0,0)
\put(140,210){\small\sc ITP-UH-01/04, TUW-04-01, YITP-SB-04-01}
\end{picture}

\abstracts{
We discuss new insights into the
quantum physics of solitons developed since 1997: why quantum
corrections to the mass $M$ and the central charge $Z$ of
solitons in supersymmetric (susy) field theories in 1+1 and 2+1
dimensions are nonvanishing, despite the fact that the 
zero-point energies of bosons and fermions seem to cancel each other,
and the central
charge is an integral of a total space derivative which naively
seems to get contributions only from regions far removed from the
soliton. Crucial are: (1) the requirement that the regularization
scheme not only makes calculations finite, but it also should
preserve (ordinary) supersymmetry, (2) the renormalization
condition that tadpoles vanish in the {\it trivial} vacuum, (3) an
anomaly in the central charge which is actually needed to
saturate the Bogomolnyi bound, (4) the influence of
the winding of classical fields on the {quantum} fields
far away from the soliton.
A new result is announced\protect\cite{Rebhan:2003bu}: for the susy
vortex solution in 2+1 dimensions, the quantum corrections to $M$
and $Z$ are both nonvanishing, but they continue to saturate the
quantum Bogomolnyi bound. Claims in the literature that no ``multiplet
shortening'' arises and hence saturation need not take place,
are shown to be incorrect because they are based on the
assumption that a second fermionic zero mode exists. We show 
however that
the latter is singular, and thus must be rejected.}

\section{Introduction}
The Casimir effect usually deals with zero point energies in the
space between plates, or between a small ball and a plate as we
heard at this conference. In certain quantum field theories one
has solitons (time-independent nonsingular  solutions of the {\it
classical} field equations in Minkowski space with {\it finite
energy}), and the quantum fluctuations around these classical
solutions also lead to zero-point energies and thus a Casimir
effect. Whereas one might perhaps argue that the Casimir effect
is only a dual formulation of the van der Waals forces between
the plates, it is much easier to compute the forces
using the Casimir approach. In quantum field theory, one
could consider two (or more) solitons, and calculate their
attraction due to the Casimir effect, but here we shall be
interested in the vacuum energy around one soliton. This is a
topic which has an enormous literature. For a review of the
situation till the 1980's, see~\cite{Rajaraman:1982is}.

Since the early 1970's when susy was discovered, it is known that
in susy  field theories the bosonic and fermionic zero point
energies {\it naively} cancel. However, the
arguments that the zero point energies should cancel are
incomplete when the background is a soliton; in fact, quantum
corrections to the soliton mass in supersymmetric field theories
are in general nonvanishing\cite{Schonfeld:1979hg,Kaul:1983yt,Imbimbo:1984nq,Rebhan:1997iv,Nastase:1998sy,Graham:1998qq,Shifman:1998zy}. 
In addition to the mass $M$,
solitons in supersymmetric theories have another quantum number,
their central charge, usually denoted by $Z$. This central charge
is classically a topological quantity: it is given by the space
integral of a total space derivative, and hence only depends on
the fields at the boundary at infinity. For a long time it was
believed that the quantum corrections to $Z$ should vanish
because all quantum corrections would have to occur far away from
the soliton where physics should be the same as if no soliton
were present. It was noticed in 1997 that this would mean the Bogomolnyi
bound\cite{Bogomolny:1976de} $M=|Z|$ would be
violated\cite{Rebhan:1997iv}. However, although the interactions
originating at the soliton can be ignored because they are short
range, the topology of these asymptotic fields reflects the
presence of a soliton somewhere in the middle. We shall show that
nonvanishing quantum corrections to the central charges of
soliton may arise either due to a new kind of
anomaly\cite{Nastase:1998sy,Shifman:1998zy} (in an odd number
of space dimensions), or due to the topology of the fields at
infinity, or perhaps due to both effects.

One way to intuitively understand this anomaly is to note that if
one uses point splitting to regulate the expression for the
central charge at the quantum level, the total derivative ceases
to be a total derivative, and its space integral ceases to vanish.
Another way, appreciated by people who are familiar with susy, is to
note that the well-known conformal-susy anomaly $\g\cdot j$ and
the well-known trace anomaly $T_{\m}^{\ \m}$ belong to a susy
multiplet of anomalies  that also contains the central charge
anomaly. If one has one of these anomalies, and one does not
break ordinary susy, the other anomalies are
inevitable\cite{Shifman:1998zy}.

In susy field theories, the corresponding superalgebra may contain
terms with central charges. For example, for the susy $N=1$ Higgs
system in 1+1 dimensions (the susy kink) the superalgebra reads 
\begin{equation}
\{Q^+,Q^+\}=H+Z,\ \ \ \{Q^-,Q^-\}=H-Z,\ \ \ \{Q^+,Q^-\}=P, \end{equation}
where $Q^{\pm}$ are the susy generators (real 2-component spinors
whose components are denoted by  + and  $-$),  $H$ is the
Hamiltonian, and $P$ the translation generator. (One could add
the Lorentz generator $J_{01}$ but we shall not need it). The
central charge generator $Z$ takes on the following form in terms
of the real scalar Higgs field $\vf$ \begin{equation}
Z=\int_{-\infty}^{\infty}dx\pa_xW(\vf), \ \ \
\pa_xW(\vf)=U(\vf)\pa_x\vf. 
\end{equation} 
The potential $V$ is related to $U$ by $V=\frac12U^2$.

In 1+1 dimensions we can also construct an $N=2$ susy model with
a kink, but then there are no quantum corrections to $M$ and
$Z$.\cite{Nastase:1998sy} In 2+1 dimensions there is an $N=2$ susy
extension of the abelian Maxwell-Higgs
model\cite{Schmidt:1992cu,Edelstein:1994bb} (one can also add a
Chern-Simon term to this system\cite{Lee:1995pm,Lee:1992yc}) and
this model we discuss below. One can truncate this $N=2$ model
down to an $N=1$ model. And in 3+1 dimensions, $N=2$ models have
susy monopoles but whether $M$ and $Z$ receive quantum
corrections in these models is an open question%
\cite{Kaul:1984bp,Imbimbo:1985mt}
(we are studying these
systems).

\section{The Nielsen-Olesen vortex with $N=2$ susy}
{\bf a. Classical model:} 
We now study well-known  Nielsen-Olesen vortex,
an abelian Maxwell-Higgs system. We consider the $N=2$
supersymmetric extension which contains in addition to the
abelian $U(1)$ gauge field $A_m$ ($m=0,1,2$)
a complex scalar $\f$ and a
complex 2-component spinor $\j$ (which form together a $N=2$
``matter multiplet'' in 2+1 dimensions), and further a complex
2-component spinor $\c$ called the gaugino and a  real scalar $N$
(the fields $A_m$, $\chi,N$ form the $N=2$ vector multiplet in
2+1 dimensions). We take the $N=2$ model instead of the $N=1$
model because we shall use {\it dimensional regularization}, and
for this scheme to preserve susy we need to find a model in higher
dimensions (in 3+1 dimensions) which is susy and which yields
upon dimensional reduction our susy vortex in 2+1 dimensions. The
simplest case is an $N=1$ model in 3+1 dimensions, whose action
reads in superspace ${\cal L}=\int d^2\q W^\a W_\a + \int
d^4\q\bar{\phi}e^{-eV}\phi+\kappa\int d^4\q V$. Dimensional reduction
(ignoring dependence on the coordinate $z$, and decomposing
$A_\m\rightarrow \{A_m$, $A_3=N$\} with $\m=0,1,2,3$ but
$m=0,1,2$) yields
the $N=2$ model in 2+1 dimensions\footnote{Our conventions are
$\eta^{\mu\nu}=(-1,+1,+1,+1)$, $\chi^{\alpha}=
\epsilon^{\alpha\beta}\chi_\beta$ and $\bar\chi^{\dot\alpha}=
\epsilon^{\dot\alpha\dot\beta}\bar\chi_{\dot\beta}$ with
$\epsilon^{\alpha\beta}=\epsilon_{\alpha\beta}=
-\epsilon^{\dot\alpha\dot\beta}=-\epsilon_{\dot\alpha\dot\beta}$
and $\epsilon^{12}=+1$. In particular we have
$\bar{\psi}_{\dot\alpha} =(\psi_\alpha)^* $ but
$\bar\psi^{\dot\alpha}=- (\psi^\alpha)^* $. Furthermore,
$\bar\sigma^\mu_{\dot\alpha\beta}=(-\mathbf 1,\vec\sigma)$ with
the usual representation for the Pauli matrices $\vec\sigma$, and
$\bar\sigma^{\mu\dot\alpha\beta}=\sigma^{\mu\beta\dot\alpha}$ with
$\sigma^{\mu\beta\dot\alpha}=(\mathbf 1,\vec\sigma)$.} 
\begin{eqnarray}
\mathcal L &=& -\4 F_{\mu\nu}^2 +\5\chi^{\.\alpha} i
\5\sigma_{\.\alpha\beta}^\mu \6_\mu \chi^\beta +\2 D^2+
(\kappa-e|\phi|^2)D\nn &&-|D_\mu \phi|^2+ \5\psi^{\.\alpha} i
\5\sigma_{\.\alpha\beta}^\mu D_\mu \psi^\beta +|F|^2 +\sqrt2 e
\left[ \phi^* \chi_\alpha \psi^\alpha +\phi \5\chi_{\.\alpha}
\5\psi^{\.\alpha} \right], \end{eqnarray} where $D_\mu=\6_\mu - ieA_\mu$
when acting on $\phi$ and $\psi$, and $F_{\mu\nu}=\6_\mu
A_\nu-\6_\nu A_\mu$. Elimination of the auxiliary field $D$
yields the scalar potential $\mathcal V=\2 D^2=\2
e^2(|\phi|^2-v^2)^2$ with $v^2\equiv {\kappa/e}$.

 The
classical vortex solution is given by 
\begin{equation} \pv = e^{in\theta}
f(r), \quad eA_+^{\mathrm V} = -i e^{i\theta}\frac{a(r)-n}{r},
\quad A_\pm^{\mathrm V} \equiv A_1^{\mathrm V} \pm i A_2^{\mathrm
V} \end{equation} or, alternatively (with $k,l=1,2$), 
\begin{equation}
\pv=\pv^1+i\pv^2=\left( \frac{x^1+ix^2}{r}\right)^n f(r), 
\quad eA_k^{\mathrm V}=\frac{\epsilon_{kl}x^l}{r}\frac{a(r)-n}{r},
\end{equation}
where $f'(r)=\frac{a}{r}f(r)$ and $a'(r)=r e^2(f(r)^2-v^2)$
with boundary conditions
\begin{eqnarray} &a(r\to\infty)=0, &f(r\to\infty)=v,\nn &a(r\to0)=n+O(r^2),
&f(r\to0)\to r^n+O(r^{n+2}). \end{eqnarray} We shall mainly discuss the
solution with $n=1$.

\smallskip
{\bf b. Quantization and renormalization:} we add a gauge fixing
term which diagonalizes the kinetic term (an $R_{\xi}$ gauge) and
from it we obtain in the usual way the corresponding
Faddeev-Popov ghost action. Setting $\f=\f_V+\eta$ and
$A_\m=A^V_\m+a_\m$, the gauge-fixing term is quadratic in quantum
fields \begin{equation}\label{Lgfix} \mathcal L_{\rm g.fix}= -\frac{1}{2\xi}
(\6_\mu a^\mu - ie\xi(\pv \eta^* - \pv^* \eta ))^2. \end{equation} The
corresponding Faddeev-Popov Lagrangian reads \begin{equation} \mathcal L_{\rm
ghost}= b \left( \6_\mu^2 - e^2 \xi \left\{ 2\,|\pv|^2 + \pv
\eta^* + \pv^* \eta  \right\} \right) c\,. \end{equation} There are no
divergences in 2+1 dimensions at the one loop  level if we use
dimensional regularization, hence one does not need
renormalization to make loops finite, but one still must account
for {\it finite renormalization}. As always in quantum field
theory, we require that the vacuum expectation value (vev) of
quantum fields vanishes, $\ll\eta\rr=0$. (If $\ll\eta\rr\neq 0$,
one must add infinitely many trees, and summing these one regains
the case that $\ll\eta\rr=0$). We have two sectors in the theory:
the trivial sector where $\phi = v$ is a solution, and the vortex
sector where the vortex solution forms the background. In the
trivial sector we set $\phi=v+\eta$, and $\frac{\kappa}{e}\equiv
v_0^2=v^2+\d v^2$, and fix $\d v^2$ such that tadpoles vanish in
the trivial vacuum: 
\begin{eqnarray}\label{sigmatadpoles}
&&\epsfbox{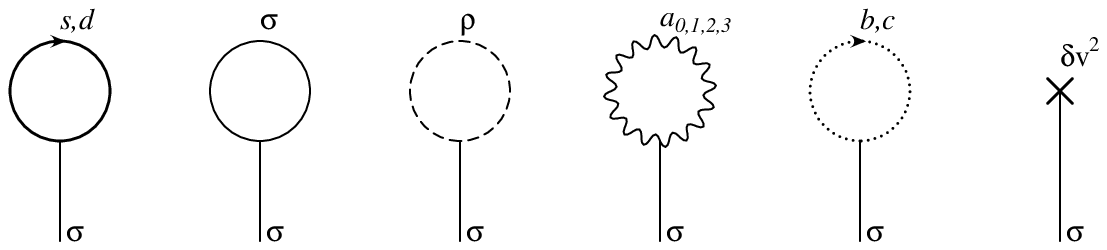}
\\
&& =(-em)\times\nonumber\\
&&\{-2 {\rm tr}{\mathbf 1_2}I(m) + \frac{3}{2}I(m) + \2 I(\xi^{\frac{1}{2}}m)
+ [3I(m)+\xi I(\xi^{\2}m)] - \xi I(\xi^{\2}m) -\delta v^2 \}.\nonumber
\end{eqnarray} 
The tadpoles consist of two fermion loops (with $s=(\j+i\c)/\sqrt2$
and $d=(\j-i\c)/\sqrt2$), the Higgs scalar $\s$, the Goldstone boson
$\rho$, the fields $a_m$ and $N$ (together they yield $a_\m$), the
ghosts $b,c$, and the counter term $\d v^2$. Then we go to the
vortex sector and set $\f=\f_V+\eta$, $A_\m=A_\m^V+a_\m$, but we
keep $v_0^2=v^2+\delta v^2$ with the $\d v^2$ which was fixed in the
trivial sector. Technically we are using a {\it minimal renormalization}
scheme. Other schemes have been discussed in~\cite{Rebhan:2002uk,%
Rebhan:2002yw}.

Note that one cannot require that tadpoles vanish in the soliton sector,
because there $\langle \eta \rangle = \phi_1(x)$ is nonvanishing,
which plays a role in the calculation of the energy densities%
\cite{Goldhaber:2001rp}.

\smallskip
{\bf c. The mass correction:} The terms in the action which are
independent of any quantum fields $(\eta,a_\m,\psi,\c)$ but which
depend on $\d v^2$ lead to a mass counter term $\D M$. (Recall
that the quantum mass of a soliton is the ``vacuum''
expectation value of the quantum
Hamiltonian). The one-loop quantum correction $M^{(1)}$ to the
mass is then given by the Casimir sum \begin{equation}\label{nine}
M^{(1)}=\pm\left(\sum\frac12\hbar \o_n -\sum\frac12 \hbar \o_n^0
\right) +\D M \end{equation} where  the contributions $\sum\frac12 \hbar
\o_n^0$ from the trivial sector vanish since we are dealing with a
susy system, while $\sum\frac12 \hbar \o_n$ sums all discrete and
continuous frequencies of the bosonic quantum fluctuations (with
a + sign) and fermionic quantum fluctuations (with a $-$ sign).

Working out the linearized field equations for the fluctuations,
one finds that the bosonic and fermionic {\it nonzero} modes have
the same solutions in 1-1 correspondence. The zero modes do not
contribute to~(\ref{nine}). This by itself does not mean that the
total sum $\sum\frac12 \hbar \o$ vanishes. If one puts the system
in a box with boundary conditions on bosons and fermions, one
finds in general spurious energy density near the
boundaries due to distortion of the various
fields\cite{Shifman:1998zy}. This spurious boundary energy must
be subtracted, and then one finds a nonvanishing remainder. One
can also work without boundaries, but in the continuum one finds
that the difference of spectral densities $\r_B^{(k)}-\r_F^{(k)}$
for the bosonic fluctuations and fermionic fluctuations is in
general nonzero, and depends on the fields at infinity. (One can
also express this difference in terms of a phase shift $\q'(k)$,
but that requires analytical formula for $\q'(k)$ which is in
general not known. We therefore developed an expression\cite{Yam} which only
depends on asymptotic values of fields in $x$-space). However,
one finds that for the susy vortex {\it all nonzero modes
cancel}. So, $M^{(1)}=\D M$, and since $\D M$ is linear in $\d
v^2$, and $\d v^2$ is finite but nonzero, one finds a finite but
nonzero value for $M^{(1)}$ (see also \cite{Vassilevich:2003xk}).

[The details are as follows
(for notation see \cite{Rebhan:2003bu}). 
The continuous part of the contributions
to the energy can be written can be written as
\begin{eqnarray}
\langle H_{\rm cont}\rangle &=& \sum (\frac12  \omega_{\rm bos}
-\frac12  \omega_{\rm ferm})
= \sum (\omega_U-\omega_V)\nonumber\\
&=& \int {d^2k\over(2\pi)^2}{d^\epsilon \ell\over(2\pi)^\epsilon}
\omega_{k,\ell}
\int d^2x [U_{\mathbf k}^\dagger U_{\mathbf k}
-V_{\mathbf k}^\dagger V_{\mathbf k}](x)
\end{eqnarray}
where $\omega_{k,\ell}=(k^2+\ell^2+m^2)^{1/2}$ and $\mathbf k$ labels
the mode functions. Using $V_{\mathbf k}=\omega_k^{-1}LU_{\mathbf k}$,
partially integrating the $V^\dagger V$ term, and using
$L^\dagger LU=\omega_k^2 U$, only a surface integral remains
which can be written as
\begin{equation}
\int d^2x (\partial_- F_+ - \partial_+ F_-)
=\int d^2x (\partial_x G_y - \partial_y G_x)=\oint d\theta G_\theta
\end{equation}
where $G_\theta=x G_y-y G_x$ has the following form
\begin{equation}
G_\theta=x^- u_1^* i D_+^{\mathrm V} u_1
+x^- \sqrt2 e \phi_{\mathrm V} u_1^* u_2 - x^+ u_2^* i \partial_- u_2
- x^+ \sqrt2 e \phi^*_{\mathrm V} u_2^* u_1.
\end{equation}
Using $x^- \partial_+ = r\partial_r + i \partial_\theta$ and
$x^+ \partial_- = r\partial_r - i \partial_\theta$
and $x^- e A_+^{\mathrm V} \to in$, we find
$\oint d\theta [u_1^* (-\partial_\theta+in)u_1+u_2^* \partial_\theta u_2]
\to 0$, because
the components $u_1$ and $u_2$ of $U$ fall off as $r^{-1/2}$ for large
$r$.
The terms with $\phi$ and $\phi^*$ cancel, and here $N=2$ susy is at
work. Far away $u_1$ tends to the free mode functions multiplied
by $e^{i n\theta}$ due to the vortex background, and
for free fields the finite (because regularized) expressions
$(u_j^* \partial_\theta u_j)$
vanish upon symmetric integration.]

\smallskip
{\bf d. The central charge correction:} Dimensional reduction of
the model in $3+1$ dimensions leads in 2+1 dimensions to the
following expression for the central charge
\begin{eqnarray}
Z=\int d^2x \left[ \e_{kl}\pa_k\xi_l+P_3\right];\ \ \xi_l=e(v^2+\d v^2)A_l-i\f^\dag D_l\f \nonumber\\
P_3=F_{0i}F_{3i}+(D_0\f)^\dag D_3\f+D_3\f^{\dag}D_0\f-i\bar{\chi}\bar{\s_0}\pa_3\c-
i\bar{\psi}\bar{\s_0}D_3\psi
\end{eqnarray}
Dimensional regularization has brought in a new term not present
in 2+1 dimensions: the term $P_3$. In the case of the kink such a
term yields a nonvanishing contribution to $Z$ which has been
interpreted to be a new anomaly. (If the extra dimension shrinks
to zero, classically $P_3$ tends to zero, but at the quantum
level $\ll P_3 \rr$ contains also an infinity due to the summing
over infinitely many modes, and ``$0\times \infty=$anomaly"). But
in odd dimensions there are no divergences and thus no anomalies,
so one expects $P_3$ to vanish. Careful study shows it indeed
vanishes. But we still need a nonzero connection for $Z$! Where
is it? The answer is rather simple: most of the loop corrections in
$\xi_l$ cancel
against the contribution from $\d v^2$ in $\xi_l$, but there is
one term left over: 
\begin{equation} Z=-i\int_0^{2\pi} d\q
\left\langle
\eta^{\dag}\frac{\pa}{\pa\q}\eta 
\right\rangle\Big|_{r=\infty}
\end{equation} 
Asymptotically
$|D_k\eta|^2\rightarrow
|\pa_r\eta|^2+\frac{1}{r^2}|(\pa_\q-in)\eta|^2$, so $\eta$
fluctuations have an extra phase $e^{in\q}$ compared with the
trivial vacuum,\footnote{One can also directly
compute $Z$ by using the mode expansion
$\eta=\sum_{m,k}R_{m,k}(r)
e^{-im\theta} e^{in\theta}$.} and this topological fact leads to a nonvanishing
$Z^{(1)}$ equal to $M^{(1)}$. Saturation holds!

\smallskip
{\bf e. Zero modes:}
It is well-known that differentiation of a classical solution with
respect to one of its parameters (such as the coordinates of its
center) yields a zero mode of the linearized field equations for the
bosonic quantum fluctuations. However, in general this zero mode solution
is only a solution of the field equations without gauge-fixing term, and
not at the same time a solution of that part of the field equations which
comes from the gauge fixing term. To obtain a solution of the complete
gauge-fixed field equations, one should add a suitable finite gauge
transformation (which is always a solution of the non-gauge-fixed
field equations). In this way one expects for the vortex two bosonic
zero modes, corresponding to the two translations in the $x$-$y$ plane.
A rotation in the $x$-$y$ plane can be undone by a gauge transformation,
so this symmetry does not produce a third bosonic zero mode.

In susy models, one can act with infinitesimal susy transformations on
the classical background, and one produces then fermionic zero modes.
For the $N=2$ vortex, one complex susy charge annihilates the bosonic
background, while the other complex susy charge generates a complex
fermionic zero mode. More in detail: the field equations for the fermionic
fluctuations are block-diagonal in two 2-dimensional spaces denoted
by $U={\psi^1 \choose \bar\chi^{\dot1}}$ and $V={\psi^2 \choose
\bar\chi^{\dot2}}$. The (iterated) field equations for $U$ and $V$
read $L^\dagger L U=(\partial_3^2-\partial_t^2)U$ and
$L L^\dagger V=(\partial_3^2-\partial_t^2)V$, where $LL^\dagger$ is a
positive definite operator which has no zero modes, but
$L^\dagger L$ has zero modes. To find these zero modes of $U$ one
must solve $LU=0$.

In the literature, it was shown that the iterated (second-order)
field equation for $\psi^1$ has two independent regular solutions, and
it was claimed that this meant that there were two independent
fermionic zero modes. However, the second solution is {\it not}
a regular solution of the original (first-order) field equation for $U$
(namely $\bar\chi^{\dot1}$ is singular at the origin). Hence there is
only one (complex) fermionic zero mode, and this implies that
$M=Z$ {\it must} hold, as we have found.

There is still a remark to be made about the relation between bosonic
and fermionic zero modes. A real bosonic zero mode (of the field
equations for $a_1$, $a_2$, Re $\eta$ and Im $\eta$) can be written
as one complex zero mode for the field equations of $B\equiv
{a_1+i a_2 \choose \eta}$. This complex bosonic zero mode is the
bosonic partner of the complex fermionic zero mode. Multiplying
this pair of complex zero modes by $i$, the result is of course again
a pair of complex zero mode solutions. However, the corresponding {\it real}
solution for $a_1$, $a_2$, and $\eta$, is {\it linearly independent}:
for example,
if the first solution corresponds to translations in the $x$ direction,
the second one corresponds to translations in the $y$ direction.
Similarly, a vortex solution with winding number $n>1$ has
$n$ complex fermionic zero modes and $2n$ real bosonic zero modes.

We conclude: there is only one (complex) fermionic zero mode (corresponding
to one pair of fermionic annihilation and creation operators), and this
gives rise to a single short multiplet at the quantum level
(a massive multiplet as short as a massless multiplet). Standard
multiplet shortening arguments\cite{Witten:1978mh} therefore do
apply and explain the preservation of the Bogomolnyi saturation
that we verified by explicit calculation at the one-loop level.
An enormous simplification in these calculations was the use
of dimensional regularization which does not need any boundary
conditions.


\begin{thebibliography}{10}
\newcommand{\enquote}[1]{``#1''}

\bibitem{Rebhan:2003bu}
A.~Rebhan, P.~van Nieuwenhuizen and R.~Wimmer, {\it Nucl. Phys.\/} {\bf B679},
  382 (2004).

\bibitem{Rajaraman:1982is}
R.~Rajaraman, {\it Solitons and Instantons\/} (North-Holland, Amsterdam, 1982).


\bibitem{Schonfeld:1979hg}
J.~F. Schonfeld, {\it Nucl. Phys.\/} {\bf B161}, 125 (1979).

\bibitem{Kaul:1983yt}
R.~K. Kaul and R.~Rajaraman, {\it Phys. Lett.\/} {\bf B131}, 357 (1983).

\bibitem{Imbimbo:1984nq}
C.~Imbimbo and S.~Mukhi, {\it Nucl. Phys.\/} {\bf B247}, 471 (1984).

\bibitem{Rebhan:1997iv}
A.~Rebhan and P.~van Nieuwenhuizen, {\it Nucl. Phys.\/} {\bf B508}, 449 (1997).

\bibitem{Nastase:1998sy}
H.~Nastase, M.~Stephanov, P.~van Nieuwenhuizen and A.~Rebhan, {\it Nucl.
  Phys.\/} {\bf B542}, 471 (1999).

\bibitem{Graham:1998qq}
N.~Graham and R.~L. Jaffe, {\it Nucl. Phys.\/} {\bf B544}, 432 (1999).

\bibitem{Shifman:1998zy}
M.~A. Shifman, A.~I. Vainshtein and M.~B. Voloshin, {\it Phys. Rev.\/} {\bf
  D59}, 045016 (1999).

\bibitem{Bogomolny:1976de}
E.~B. Bogomolnyi, {\it Sov. J. Nucl. Phys.\/} {\bf 24}, 449 (1976).

\bibitem{Schmidt:1992cu}
J.~R. Schmidt, {\it Phys. Rev.\/} {\bf D46}, 1839 (1992).

\bibitem{Edelstein:1994bb}
J.~Edelstein, C.~N{\'u\~n}ez and F.~Schaposnik, {\it Phys. Lett.\/} {\bf B329},
  39 (1994).

\bibitem{Lee:1995pm}
B.-H. Lee and H.~Min, {\it Phys. Rev.\/} {\bf D51}, 4458 (1995).

\bibitem{Lee:1992yc}
B.-H. Lee, C.-k. Lee and H.~Min, {\it Phys. Rev.\/} {\bf D45}, 4588 (1992).

\bibitem{Kaul:1984bp}
R.~K. Kaul, {\it Phys. Lett.\/} {\bf B143}, 427 (1984).

\bibitem{Imbimbo:1985mt}
C.~Imbimbo and S.~Mukhi, {\it Nucl. Phys.\/} {\bf B249}, 143 (1985).

\bibitem{Rebhan:2002uk}
A.~Rebhan, P.~van Nieuwenhuizen and R.~Wimmer, {\it New J. Phys.\/} {\bf 4}, 31
  (2002).

\bibitem{Rebhan:2002yw}
A.~Rebhan, P.~van Nieuwenhuizen and R.~Wimmer, {\it Nucl. Phys.\/} {\bf B648},
  174 (2003).

\bibitem{Goldhaber:2001rp}
A.~S. Goldhaber, A.~Litvintsev and P.~van Nieuwenhuizen, {\it Phys. Rev.\/}
  {\bf D67}, 105021 (2003).

\bibitem{Yam}
A.~S. Goldhaber, A.~Rebhan, P.~van Nieuwenhuizen and R.~Wimmer,
  \enquote{Quantum corrections to mass and central charge of supersymmetric
  solitons}, contribution to the Hidenaga Yamagishi commemorative volume of
  Physics Reports, edited by E. Witten and I. Zahed.


\bibitem{Vassilevich:2003xk}
D.~V. Vassilevich, {\it Phys. Rev.\/} {\bf D68}, 045005 (2003).

\bibitem{Witten:1978mh}
E.~Witten and D.~Olive, {\it Phys. Lett.\/} {\bf B78}, 97 (1978).

\end{thebibliography}

\end{document}